\documentclass[12pt]{article}
\usepackage{amsmath}
\usepackage{amssymb}
\usepackage{graphicx}
\usepackage{float}
\usepackage{indentfirst}
\usepackage{mathrsfs}
\usepackage{dsfont}

\usepackage{setspace}

\usepackage[labelfont=bf,labelsep=period]{caption}

\usepackage[numbers,sort&compress]{natbib}

\newtheorem{theorem}{Theorem}

\newtheorem{corollary}{Corollary}
\newtheorem{lemma}{Lemma}
\newtheorem{observation}{Observation}
\newtheorem{conjecture}{Conjecture}
\title{A Necessary and Sufficient Criterion for the Separability of Quantum State}

\author
{Jun-Li Li$^{1,3}$ and Cong-Feng Qiao$^{1,2,3\ast}$\\ [0.2cm]
\normalsize{$^1$Department of Physics, University of the Chinese Academy of Sciences,}\\
\normalsize{YuQuan Road 19A, Beijing 100049, China}\\[2pt]
\normalsize{$^2$ Department of Physics \& Astronomy, York University, Toronto, ON M3J 1P3, Canada}\\[2pt]
\normalsize{$^3$Key Laboratory of Vacuum Physics, University of Chinese Academy of Sciences}\\[3mm]
\normalsize{$^\ast$ To whom correspondence should be addressed; E-mail: qiaocf@ucas.ac.cn.}
}

\date{}
\begin{document}
\baselineskip24pt \maketitle
\vspace{0.5cm}
\begin{abstract} \doublespacing
Quantum entanglement has been regarded as one of the key physical resources in quantum information sciences. However, the determination of whether a mixed state is entangled or not is generally a hard issue, even for the bipartite system. In this work we propose an operational necessary and sufficient criterion for the separability of an arbitrary bipartite mixed state, by virtue of the multiplicative Horn's problem. The work follows the work initiated by Horodecki {\it et. al.} and uses the Bloch vector representation introduced to the separability problem by J. De Vicente. In our criterion, a complete and finite set of inequalities to determine the separability of compound system is obtained, which may be viewed as trade-off relations between the quantumness of subsystems. We apply the obtained result to explicit examples, e.g. the separable decomposition of arbitrary dimension Werner state and isotropic state.
\end{abstract}

\newpage

\section{Introduction}

Entanglement is a ubiquitous feature of quantum system and key element in quantum information processing, whereas yet far from fully understood \cite{NC-Book}. A fundamental problem in the study of entanglement is the determination of the separability of quantum states. For pure state, the entangled states are those that cannot be expressed as the product of the subsystems, e.g. we say a bipartite pure state of $A$ and $B$ is entangled if it cannot be expressed in the product form like
\begin{eqnarray}
|\psi\rangle_{AB} = |\phi\rangle_{A} \otimes |\varphi\rangle_{B} \; .
\end{eqnarray}
For the mixed state of a compound system, we say it is entangled if it cannot be written as a convex combination of product states. For example, a bipartite mixed state is separable (i.e. classically correlated \cite{Mixed-ent}) whenever it can be expressed as
\begin{eqnarray}
\rho_{AB} = \sum_{i=1}^L p_i \rho_i^{(A)} \otimes \rho_i^{(B)} \; . \label{bipartite-sum}
\end{eqnarray}
Here, $\rho_i^{(A,B)}$ are local density matrices of particles $A$ and $B$; $p_i>0$ and $\sum_{i=1}^L p_i = 1$. The entanglement (non-separability) criterion for pure state is clear, by virtue of Schmidt or high order singular value decomposition for any-number-partite system \cite{LU-Multi}. However, none of the existing criteria for the separability of finite dimensional mixed states are satisfactory by far. They are generally either sufficient and necessary, but not practically usable; or easy to use, but only necessary (or only sufficient) \cite{Geometry-State-Book}.

Over the past decades, one remarkable progress towards the operational characterization of a separable mixed state, the positive partial transposition (PPT) criterion \cite{PPT-criterion}, was achieved by Peres twenty years ago. This separability criterion is only necessary and sufficient for $2\times 2$ and $2\times 3$ systems, rather than for arbitrary higher dimensional systems \cite{H-criterion}. Though couple of necessary and sufficient criteria were developed afterwards \cite{H-criterion, Bell-criterion, Wu-NC-criterion}, they are generally difficult to handle in practice, or only applicable to low-rank density matrices \cite{low-rank-criterion}. With the dimension growing, the separability problem of a compound system tends to be NP-hard, even in the bipartite situation \cite{NP-hard}. Recent investigations mostly focus on the necessary or sufficient conditions of witnessing entanglement or separability. The computable cross-norm or realignment (CCNR) criterion \cite{cross-norm, realignement-criterion} and local uncertainty relations (LURs) \cite{LURs} are proposed to detect entanglement. By virtue of the Bloch representation, separability criterion had been successfully formulated in matrix norms, which was found to be related to the CCNR criterion \cite{Bloch-criterion}. The optimization of entanglement witness observables may stand as a separability criterion \cite{Witness-SV}. For recent development, readers may refer to Refs.\cite{CM-CC-LU, Z-R, Recent-develop2} and more comprehensive reviews \cite{Review-1, Review-2}. It should be noted that even restricted to necessary or sufficient criterion, the corresponding inequalities tend to be an ever growing family. Therefore, to find a complete and finite set of inequalities to determine the separability of mixed states is theoretically important and practically necessary.

In this work, we present an applicable criterion for the separability of bipartite mixed state through exploring the multiplicative Horn's problem \cite{Non-invertible}. By expressing the quantum state in Bloch representation, the problem of factorizing a mixed state into sum of product states is transformed to the task of decomposing a matrix into the product of two other matrices. We find that the solution to the multiplicative Horn's problem yields a complete and finite set of inequalities, a new criterion, which in practice provides a systematic procedure for the decomposition of separable mixed states. Relations between this new criterion and other related ones are investigated through concrete examples, including the separable decomposition of arbitrary dimensional Werner and isotropic states. Results manifest that the criterion raised in this work is to our knowledge the most applicable one at present in determining the separability of entangled quantum states.

\section{The separability of bipartite mixed states}
\subsection{The Bloch representation of quantum state}

A quantum state in $N$-dimensional Hilbert space may be expressed as \cite{Bloch-vector}
\begin{eqnarray}
\rho = \frac{1}{N}\mathds{1} + \frac{1}{2}  \sum_{\mu=1}^{N^2-1} r_{\mu} \lambda_{\mu} \; ,
\end{eqnarray}
where the real coefficients $r_{\mu} = \langle \lambda_{\mu} \rangle = \mathrm{Tr}[\rho\lambda_{\mu}]$, and $\lambda_{\mu}$ are the $N^2-1$ generators of SU($N$) group. The $N^2-1$ dimensional vector $\vec{r} = \{r_1,\cdots, r_{N^2-1}\}^{\mathrm{T}}$ is called Bloch vector (or coherent vector) of the density matrix $\rho$, where the superscript $\mathrm{T}$ means the transposition. As the density matrix must be positive semidefinite and normalized, the vector $\vec{r}$ subjects to a set of constraints \cite{Positivity-Bloch-a,Positivity-Bloch-b}, among which $|\vec{r}\,|\leq \sqrt{2(N-1)/N}$ is imposed by the condition $\mathrm{Tr}[\rho^2]\leq 1$ with the vector norm defined as $|\vec{r}\,| \equiv \sqrt{\vec{r}\cdot \vec{r}}$. Similarly, any bipartite state of dimensions $N\times M$ in the Bloch representation can be expressed as
\begin{eqnarray}
\rho_{AB} & = & \frac{1}{NM}  \mathds{1} \otimes \mathds{1} + \frac{1}{2M} \vec{a}\cdot \vec{\lambda}\otimes \mathds{1} + \frac{1}{2N} \mathds{1} \otimes \vec{b} \cdot \vec{\sigma} + \frac{1}{4} \sum_{\mu=1}^{N^2-1} \sum_{\nu=1}^{M^2-1} \mathcal{T}_{\mu\nu} \lambda_{\mu} \otimes \sigma_{\nu} \; . \label{bipartite-bloch}
\end{eqnarray}
Here $a_{\mu} = \mathrm{Tr}[\rho_{AB} (\lambda_{\mu}\otimes \mathds{1})]$, $b_{\nu} = \mathrm{Tr}[ \rho_{AB} (\mathds{1} \otimes \sigma_{\nu})]$, $\mathcal{T}_{\mu\nu} = \mathrm{Tr}[\rho_{AB}(\lambda_{\mu} \otimes \sigma_{\nu})]$, and $\sigma_{\nu}$ are generators of SU($M$). Reformulating the right hand side of Eq. (\ref{bipartite-sum}) in term of Bloch representation of $\rho^{(A)}_i = \displaystyle \frac{1}{N}\mathds{1} + \frac{1}{2}  \vec{r}_i \cdot \vec{\lambda}$, $\rho^{(B)}_j = \displaystyle \frac{1}{M} \mathds{1} + \frac{1}{2} \vec{s}_j \cdot \vec{\sigma}$, and comparing with Eq. (\ref{bipartite-bloch}), the necessary and sufficient condition of separability then turns to
\begin{eqnarray}
\sum_i p_i\vec{r}_i = \vec{a} \; , \; \sum_j p_j \vec{s}_j = \vec{b} \; , \;
\sum_{k=1}^{n} p_k \vec{r}_k \vec{s}_k^{\,\mathrm{T}} = \mathcal{T} \; , \label{General-form-density}
\end{eqnarray}
where the subscripts in $\vec{r}_i$, $\vec{s}_j$ label different Bloch vectors rather the components of them, and the correlation matrix $\mathcal{T}$ has the matrix elements of $\mathcal{T}_{\mu\nu}$.

The reduced density matrices of $A$ and $B$ can be derived from  Eq. (\ref{bipartite-bloch}) by partial trace
\begin{align}
\rho_A = \mathrm{Tr}_B[\rho_{AB}] = \frac{1}{N} \mathds{1} + \frac{1}{2}\vec{a} \cdot \vec{\lambda}\; , \;
\rho_B = \mathrm{Tr}_A[\rho_{AB}] = \frac{1}{M} \mathds{1} + \frac{1}{2}\vec{b} \cdot \vec{\sigma} \; .
\end{align}
Here the local ranks are $\mathrm{rank}(\rho_A) = n$ and $\mathrm{rank}(\rho_{B})=m$, where $n$ and $m$ may be non-full local ranks of $n<N$ and $m<M$. We then have the following observation (see \cite{Sep-APP} for details):
\begin{observation}
All $N\times M$ mixed states with local ranks $n<N$ and $m<M$ are either reducible to $n\times m$ states with full local ranks or entangled. \label{observ-0}
\end{observation}
According to Observation \ref{observ-0}, we need only consider the separability of mixed states with full local ranks. The full local rank state could be transformed to a normal form with maximally mixed subsystems \cite{Normal-Form}, i.e.
\begin{equation}
\rho_{AB} \to \widetilde{\rho}_{AB} = \frac{1}{NM}\mathds{1} + \frac{1}{4} \sum_{\mu,\nu=1}^{N^2-1,M^2-1} \widetilde{\mathcal{T}}_{\mu\nu} \lambda_{\mu} \otimes \sigma_{\nu} \;.
\end{equation}
Note that in the literature there are studies about the normal form in the separability problem \cite{separa-normal-form, normal-forms-criteria}. Hereafter, the quantum states $\rho_{AB}$ are implied to be in their normal form, and we have
\begin{observation}
Let $\vec{r}_i$ and $\vec{s}_j$ be Bloch vectors of density matrices and $\vec{p} = (p_1,p_2,\cdots, p_L)^{\mathrm{T}}$, we may define two matrices $M_{rp} \equiv M_{r}D_{p}^{\frac{1}{2}}$ and $M_{sp} \equiv M_{s}D_{p}^{\frac{1}{2}}$, where $M_r = \{\vec{r}_1, \vec{r}_2, \cdots, \vec{r}_L\}$, $M_s = \{ \vec{s}_1, \vec{s}_2, \cdots, \vec{s}_L\}$, and $D_{p} = \mathrm{diag}\{p_1, p_2, \cdots, p_L\}$ with $0< p_i\leq 1$ and $\sum_{i=1}^L p_i = 1$. The state $\rho_{AB}$ is separable if and only if there exist a number $L$ such that $\mathcal{T} = M_{rp} M_{sp}^{\,\mathrm{T}}$ with $M_r \vec{p} = 0$ and $M_s\vec{p} = 0$. \label{observ-1}
\end{observation}

\subsection{The criterion of separability}

For arbitrary bipartite quantum state $\rho_{AB}$ in normal form, let $\mathcal{T}$ be its correlation matrix, and $M_{rp}$ and $M_{sp}$ be defined in Observation \ref{observ-1}, the decomposition $\mathcal{T}= M_{rp} M_{sp}^{\,\mathrm{T}}$ then can be obtained via the following theorem:
\begin{theorem}
A real matrix $\mathcal{T}$ can be decomposed as $\mathcal{T} =M_{rp}M_{sp}^{\mathrm{T}}$ if and only if $M_{rp}$, $M_{sp}$, and $D_{\tau}$ have the following relation
\begin{eqnarray}
M_{rp} & = & (\vec{u}_1, \cdots, \vec{u}_L) X D_{\alpha} Q^{(1)} \; , \label{Mrp-UXDQ}\\
M_{sp} & = & (\vec{v}_1, \cdots, \vec{v}_L) Y D_{\beta}  Q^{(2)} \; , \label{Msp-UYDQ}\\
D_{\tau} & = & XD_{\alpha} Q^{(1)} Q^{(2)\mathrm{T}} D_{\beta} Y^{\mathrm{T}} \; . \label{D-t-a-b}
\end{eqnarray}
Here, $\vec{u}_i$ and $\vec{v}_j$ are the left and right singular vectors of $\mathcal{T} = \sum_{i=1}^L \tau_i \vec{u}_i\vec{v}_i^{\,\mathrm{T}}$ with singular values of $\tau_i$; $X, Y, Q^{(1)}, Q^{(2)} \in \mathrm{SO}(L)$; $D_{\alpha} = \mathrm{diag}\{\alpha_1, \cdots, \alpha_L\}$ and $D_{\beta} = \mathrm{diag}\{\beta_1, \cdots, \beta_L\}$ are singular values of $M_{rp}$ and $M_{sp}$; $D_{\tau} = \mathrm{diag}\{\tau_1, \cdots, \tau_L\}$, with $L> \mathrm{rank}(\mathcal{T})=l$; all the singular values are arranged in descending order.
\label{Theorem-main1}
\end{theorem}

\noindent{\bf Proof:} The if part is quite straightforward
\begin{align}
M_{rp}M_{sp}^{\mathrm{T}} & = (\vec{u}_1, \cdots, \vec{u}_L) X D_{\alpha} Q^{(1)} Q^{(2)\mathrm{T}} D_{\beta} Y^{\mathrm{T}}
\begin{pmatrix}
\vec{v}_1^{\,\mathrm{T}} \\
\vdots \\
\vec{v}_L^{\,\mathrm{T}}
\end{pmatrix} \nonumber \\ &=  (\vec{u}_1, \cdots, \vec{u}_L) D_{\tau}
\begin{pmatrix}
\vec{v}_1^{\,\mathrm{T}} \\
\vdots \\
\vec{v}_L^{\,\mathrm{T}}
\end{pmatrix} = \mathcal{T} \; .
\end{align}
For the only if part, suppose the singular value decompositions of $M_{rp}$ and $M_{sp}$ are
\begin{align}
M_{rp} = (\vec{u}\,'_{\!1}, \cdots, \vec{u}\,'_{\!L}) D_{\alpha} Q \; ,\; M_{sp} = (\vec{v}\,'_{\!1}, \cdots, \vec{v}\,'_{\!L}) D_{\beta} Q' \; ,
\end{align}
then we have
\begin{equation}
M_{rp}M_{sp}^{\mathrm{T}} = (\vec{u}\,'_{\!1}, \cdots, \vec{u}\,'_{\!L}) D_{\alpha}QQ'^{\,\mathrm{T}} D_{\beta}
\begin{pmatrix}
\vec{v}\,'^{\,\mathrm{T}}_{\!1}\\
\vdots \\
\vec{v}\,'^{\,\mathrm{T}}_{\!L}
\end{pmatrix} \; .
\end{equation}
Because $\mathcal{T} = M_{rp}M_{sp}^{\mathrm{T}}$, the singular value decomposition of the matrix $D_{\alpha}QQ'^{\,\mathrm{T}}D_{\beta}$ is $D_{\alpha}QQ'^{\,\mathrm{T}}D_{\beta} = X^{\mathrm{T}} D_{\tau}Y$,
where $D_{\tau}$ contains the first $L$ singular values of $\mathcal{T}$. We have
\begin{equation}
M_{rp}M_{sp}^{\mathrm{T}} = (\vec{u}\,'_{\!1}, \cdots, \vec{u}\,'_{\!L}) X^{\mathrm{T}} D_{\tau}Y
\begin{pmatrix}
\vec{v}\,'^{\,\mathrm{T}}_{\!1} \\
\vdots \\
\vec{v}\,'^{\,\mathrm{T}}_{\!L}
\end{pmatrix}\;
\end{equation}
with $(\vec{u}\,'_{\!1}, \cdots, \vec{u}\,'_{\!L}) X^{\mathrm{T}} = (\vec{u}_1, \cdots, \vec{u}_L) $ and $(\vec{v}\,'_{\!1}, \cdots, \vec{v}\,'_{\!L}) Y^{\mathrm{T}} = (\vec{v}_1, \cdots, \vec{v}_L) $. Q.E.D.

Observation \ref{observ-1} turns the separable problem of a compound system to the question of how to decompose the correlation matrix into a product of other two nontrivial matrices, i.e. $\mathcal{T} = M_{rp} M_{sp}^{\mathrm{T}}$, with constraints $M_r \vec{p} = 0$ and $M_s\vec{p} = 0$. Theorem \ref{Theorem-main1} further gives the decomposition conditions, that is: 1. The left singular vectors of $M_{rp}$ and $M_{sp}$ agree with the left and right singular vectors of $\mathcal{T}$ (i.e. Eqs. (\ref{Mrp-UXDQ}, \ref{Msp-UYDQ})); 2. The right singular vectors of $M_{rp}$ and $M_{sp}$, and their singular values satisfy Eq. (\ref{D-t-a-b}). For condition 1, we may rotate the orthogonal bases of particle $A$ and $B$ to $\{\vec{u}_i\}$ and $\{\vec{v}_i\}$ respectively, where $\mathcal{T}$ becomes a diagonal matrix. While for condition 2, we need to solve Eq. (\ref{D-t-a-b}), which makes the singular values of matrices $\mathcal{T}$, $M_{rp}$, and $M_{sp}$ correlated.

Before proceeding to the Eq. (\ref{D-t-a-b}), two prerequisite lemmas are necessary. Let $I$, $J$, $K$ be certain subsets of natural numbers $\{1,\cdots, n\}$ with the same cardinality $r$, i.e., $I = \{i_1, i_2, \cdots, i_r\}$, $J = \{j_1, j_2, \cdots, j_r\}$, and $K = \{k_1, k_2, \cdots, k_r\}$, where the elements are arranged in increasing order so that $i_r> i_{r-1}> \cdots >i_1$;  $j_r> j_{r-1} > \cdots >j_1$, and $k_r> k_{r-1} > \cdots >k_1$. Define $\mathcal{F}(I) \equiv (i_r-r, i_{r-1}-(r-1),\cdots, i_1-1)$, and let the triplet $(\lambda,\mu,\nu) = (\mathcal{F}(I),\mathcal{F}(J),\mathcal{F}(K))$, then we are legitimate to introduce a triple set $T_r^n = \{(I, J, K)\}$ defined as:
\begin{lemma}
A triplet $(I,J,K)$ is in $T_r^n$ if and only if the corresponding triplet $(\lambda,\mu,\nu)$ occurs as eigenvalues of the triple of $r$ by $r$ Hermitian matrices, with the third to be the sum of the first two.
\end{lemma}
This lemma appears as Theorem 2 of Ref. \cite{Review-Fulton}, where the practical methods on how to generate $T_r^n$ were also discussed, i.e., via the Horn's inductive procedure or Littlewood-Richardson coefficients.
\begin{lemma}
A triplet $(\{a_i\}$, $\{b_i\}$,$\{c_i\})$ occurs as singular values of $n$ by $n$ real matrices $A$, $B$, and $C(C = AB)$ if and only if
\begin{eqnarray}
\prod_{k\in K} c_k \leq \prod_{i\in I} a_i \cdot \prod_{j\in J} b_j \label{semidefinite}
\end{eqnarray}
for all $(I,J,K)$ in $T_r^n$ and all $r<n$. \label{lemma-triple}
\end{lemma}
This is known as the multiplicative Horn's problem, see theorem 16 of Ref. \cite{Review-Fulton} for details. Historically, the multiplicative Horm's problem first appeared as the Thompson's conjecture \cite{Thompson-conj}, and later was found can be solved for invertible matrices \cite{Random-walks}. It was found to be true for real matrices \cite{Real-product}, and even extendable to the case of non-invertible matrices recently \cite{Non-invertible} (see Appendix for a brief review of the proof).

The decomposition of Eq. (\ref{D-t-a-b}) can be realized through the following theorem:
\begin{theorem}
There exists a real orthogonal matrix $Q$ such that $D_{\alpha}QD_{\beta}$ has the singular values of $D_{\tau}$, if and only if the following is satisfied
\begin{equation}
\prod_{k\in K} \tau_k \leq \prod_{i\in I} \alpha_i \prod_{j\in J} \beta_j \; , \label{horn-singular-values}
\end{equation}
for all $(I,J,K)\in T_r^L$ and $r<L$. \label{Theorem-main2}
\end{theorem}

Theorem \ref{Theorem-main2} is a direct application of Lemma \ref{lemma-triple}. For a given bipartite state whose correlation matrix $\mathcal{T}$ is known, Eq. (\ref{horn-singular-values}) applies to all possible singular values of the matrices decomposed from $\mathcal{T}$, and behaves as trade-off relations among them. The singular values of $M_{rp}$ and $M_{sp}$ are determined by their column vectors, i.e. $\vec{r}_i$ and $\vec{s}_i$, whose norms relate to the mixedness (or quantumness) of the subsystems, i.e. $\rho_i^{(A)}$ and $\rho_{i}^{(B)}$. Large $\tau_i$ implies large $\alpha_i$ or $\beta_i$ or both. When column vectors $\vec{r}_i$ and $\vec{s}_i$ surpass the Bloch vectors of density matrices in lengths, the quantum state $\mathcal{T}$ is entangled. The quantum state is separable only when the two factor matrices are composed of Bloch vectors of physical states.

In the following we demonstrate our method in bipartite quantum system as an example. For more systematic and detailed applications, readers may refer to Ref. \cite{Sep-APP}. It should be noted that theorems \ref{Theorem-main1} and \ref{Theorem-main2} are also suitable to the bi-separability of arbitrary multipartite states, and hence the method presented here is also applicable to the multi-separability problem, due to the reason that the Bloch representation generally turns the sum decomposition problem into a product decomposition one.

\subsection{Applications}

In Bloch representation of quantum state, we have the following two observations:
\begin{observation}
If $\vec{r}$ is a Bloch vector of a density matrix, then the $\vec{r}\,'$, whose components $r_{\mu}' = -r_{\mu}$ corresponding to those SU$(N)$ generators satisfying $\lambda_{\mu}^{\mathrm{T}} = -\lambda_{\mu}$, is also a Bloch vector of a density matrix.  \label{observ-2}
\end{observation}
\begin{observation}
If the norm of a Bloch vector with dimension $(N^2-1)$ satisfies $|\vec{r}\,|^2 \leq \frac{2}{N(N-1)}$, then $\vec{r}\,' = P\vec{r}$ is also a Bloch vector for arbitrary rotation $P \in$ SO$(N^2-1)$. \label{observ-3}
\end{observation}
Here, the Observation \ref{observ-2} is established due to the fact that the transposition of a positive semidefinite Hermitian matrix keeps on being positive semidefinite, while the Observation \ref{observ-3} is just a corollary of Eq. (11) in Ref. \cite{Bloch-criterion} (or see Ref. \cite{inner-circle}). In the following, we demonstrate how the criterion works through concrete examples.

\subsubsection{ Example I: The relationship between Vicente's criterion \cite{Bloch-criterion} and ours}

A subset inequalities of Eq. (\ref{horn-singular-values}) goes as follows (see theorem 3.3.4 of Ref. \cite{Horn-topics}):
\begin{eqnarray}
\prod_{i=1}^k \tau_{i} \leq \prod_{i=1}^k \alpha_i \beta_i \; , \; k=1,2,\cdots \; . \label{Major-ABC}
\end{eqnarray}
Employing Ky Fan matrix norm $||\mathcal{T}||_{\mathrm{KF}} = \sum_{i=1}^{L} \tau_i$ and Schwarz inequality $\sum_i \alpha_i\beta_i \leq (\sum_i\alpha_i^2)^{1/2}(\sum_i\beta_i^2)^{1/2}$, we have:
\begin{corollary}
The average square norms of the local states' Bloch vectors are lower bounded by Ky Fan norm of the correlation matrix
\begin{eqnarray}
\left(\sum_{i=1}^L p_i |\vec{r}_i|^2\right) \left(\sum_{j=1}^L p_j|\vec{s}_j|^2 \right) \geq ||\mathcal{T}||_{\mathrm{KF}}^2\; . \label{corollary-KF}
\end{eqnarray} \label{corollary-KyFan}
\end{corollary}

\noindent{\bf Proof:} Eq. (\ref{Major-ABC}) leads to (see Corollary 3.3.10 of Ref. \cite{Horn-topics})
\begin{equation}
\sum_{i=1}^{k} \tau_i \leq \sum_{i=1}^k \alpha_i\beta_i\; , \; k \in \{1,\cdots, L\} \; .
\end{equation}
The Ky Fan norm of $\mathcal{T}$ is
\begin{equation}
||\mathcal{T}||_{\mathrm{KF}} = \sum_i \tau_i \leq \sum_{i} \alpha_i\beta_i \leq (\sum_i \alpha_i^2)^{\frac{1}{2}}(\sum_i \beta_i^2)^{\frac{1}{2}} \; .
\end{equation}
The Frobennius norm for real matrices are $||M||_2 \equiv \mathrm{Tr}[M^{\mathrm{T}}M]^{\frac{1}{2}}$, so we have
\begin{align}
\sum_{i}\alpha_i^2 = \mathrm{Tr}[M_{rp}^{\mathrm{T}}M_{rp}] = \sum_i p_i |\vec{r}_i|^2 \; , \;
\sum_{i}\beta_i^2 = \mathrm{Tr}[M_{sp}^{\mathrm{T}}M_{sp}] =\sum_i p_i |\vec{s}_i|^2 \; .
\end{align}
Q.E.D.

Because $|\vec{r}_i|^2 \leq \frac{2(N-1)}{N}$ and $|\vec{s}_i|^2 \leq \frac{2(M-1)}{M}$, we have
\begin{align}
\sum_{i=1}^{L} p_i |\vec{r}_i|^2 \leq \frac{2(N-1)}{N} \; , \; \sum_{i=1}^L p_i |\vec{s}_i|^2 \leq \frac{2(M-1)}{M} \; . \label{Average-norm}
\end{align}
Taking Eq. (\ref{Average-norm}) into Corollary \ref{corollary-KyFan}, Theorem 1 of Ref. \cite{Bloch-criterion} is arrived. On the other hand, from Observation \ref{observ-3} we have the following:
\begin{corollary}
If $ ||\mathcal{T}||_{\mathrm{KF}} \leq \displaystyle \frac{2}{\sqrt{MN(M-1)(N-1)}}$, the quantum state $\mathcal{T}$ is separable. \label{corollary-in}
\end{corollary}

\noindent{\bf Proof:} Suppose that $\mathcal{T} =\sum_{i=1}^l \tau_i \vec{u}_i\vec{v}_i^{\mathrm{T}}$ with rank$(T)=l\geq 1$, when working in the bases of $\vec{u}_i$ and $\vec{v}_i$, we may construct the following matrix equation
\begin{equation}
\begin{pmatrix}
\tau_1 & 0 & \cdots & 0 & 0 \\
0 & \tau_2 & \cdots & 0 & 0\\
\vdots & \vdots & \ddots & \vdots & \vdots \\
0 & 0 & \cdots  & \tau_l & 0 \\
0 & 0 & \cdots & 0 & 0
\end{pmatrix} =
\begin{pmatrix}
\alpha_1 & 0 & \cdots & 0 & 0\\
0 & \alpha_2 & \cdots & 0 & 0\\
\vdots & \vdots & \ddots & \vdots & \vdots\\
0 & 0 & \cdots  & \alpha_l & 0 \\
0 & 0 & \cdots &  0 & 0
\end{pmatrix} QQ^{\mathrm{T}}
\begin{pmatrix}
\beta_1 & 0 & \cdots & 0 & 0\\
0 & \beta_2 & \cdots & 0 & 0\\
\vdots & \vdots & \ddots & \vdots & \vdots\\
0 & 0 & \cdots  & \beta_l & 0 \\
0 & 0 & \cdots &  0 & 0
\end{pmatrix} \; . \label{minimum-cosntruct}
\end{equation}
Here, $Q\in \mathrm{SO}(l+1)$ with elements in the last row being $Q_{(l+1)j} = \sqrt{p_j}$; $p_j\geq 0$, and $\sum_{j=1}^{l+1} p_j = 1$. Choosing $\alpha_i = (\frac{2}{N(N-1)})^{\frac{1}{2}} \sqrt{\kappa_i}$, $\beta_i = (\frac{2}{M(M-1)})^{\frac{1}{2}}\sqrt{\kappa_i}$, and $\kappa_i = \tau_i \frac{\sqrt{N(N-1)M(M-1)}}{2}$, we have
\begin{equation}
\tau_i = \alpha_i\beta_i\; , \; \mathcal{K} \equiv \sum_{i=1}^l \kappa_i = \frac{\sqrt{N(N-1)M(M-1)}}{2}  \sum_{j=1}^l \tau_i\leq 1\;.
\end{equation}
Comparing Eq. (\ref{minimum-cosntruct}) with $\mathcal{T} = M_{rp}M_{sp}^{\mathrm{T}}$, we can get the Bloch vectors $\vec{r}_j$ and $\vec{s}_j$
\begin{align}
\sqrt{p_j} \,\vec{r}_j & = (\alpha_1 Q_{1j}, \alpha_2 Q_{2j}, \cdots, \alpha_l Q_{lj})^{\mathrm{T}} \; , \label{rp-r} \\
\sqrt{p_j} \, \vec{s}_j & = (\beta_1 Q_{1j}, \beta_2 Q_{2i}, \cdots, \beta_l Q_{lj})^{\mathrm{T}}\; ,\label{sp-s}
\end{align}
where $j \in \{1,\cdots, l+1\}$ and the norms are
\begin{align}
p_{j}|\vec{r}_j|^2 & = \sum_{i=1}^l \alpha^2_i Q_{ij}^2 = \frac{2}{N(N-1)} \sum_{i=1}^l \kappa_i Q^2_{ij} \; , \label{norm-rp} \\
p_{j}|\vec{s}_j|^2 & = \sum_{i=1}^l \beta_i Q_{ij}^2 = \frac{2}{M(M-1)} \sum_{i=1}^l \kappa_i Q^2_{ij}\; . \label{norm-sp}
\end{align}
We may set the probability distribution $p_j$ to be $p_j = \frac{1}{\mathcal{K}} \sum_{i=1}^l \kappa_iQ^2_{ij}$. Then replacing the $p_j$ in Eqs. (\ref{norm-rp}, \ref{norm-sp}), we have
\begin{equation}
|\vec{r}_j|^2 = \frac{2\mathcal{K}}{N(N-1)} \leq \frac{2}{N(N-1)} \; , \; |\vec{s}_j|^2 = \frac{2\mathcal{K}}{M(M-1)} \leq \frac{2}{M(M-1)} \; .
\end{equation}
According to Observation \ref{observ-3}, the Corollary \ref{corollary-in} is then established. Q.E.D.

Corollary \ref{corollary-in} agrees with the Proposition 3 of Ref. \cite{Bloch-criterion} where $M$ and $N$ are dimensions of the subsystems. Here, in proof of Corollary \ref{corollary-in}, explicit separable decomposition of $\mathcal{T}$ into $M_{rp}$ and $M_{sp}$ with $M_{r}\vec{p}=M_{s}\vec{p} = 0$ also exhibits.

\subsubsection{ Example II: The relation with PPT \cite{PPT-criterion} scheme}

The partial transposition of a bipartite density matrix corresponds to the sign flips of columns or rows (not both) of $\mathcal{T}$, whose indices are that of skew symmetric generators, i.e., $\lambda_{\mu}^{\mathrm{T}}=-\lambda_{\mu}$. The Observation \ref{observ-2} implies that the PPT criterion is necessary for separability. Conversely, the positivity of partially transposed density matrix generally does not imply separability, that is, PPT is not sufficient. However, for qubit-qubit system, calculation indicates that the PPT of density operators gives $0\leq \sum_i \tau_i \leq 1$ by means of the technique introduced in Ref. \cite{two-qubit-bloch} (see the Appendix). As $1\leq \frac{2}{2(2-1)}$, according to the Corollary \ref{corollary-in}, PPT also tells separability. Therefore it is a necessary and sufficient condition for qubit-qubit system, which agrees with the conclusionn proved by other method \cite{H-criterion}.

\subsubsection{Example III: For generalized Werner state and isotropic state}

\noindent{\bf The relation between the Werner state and the isotropic state}

The generalized Werner state and isotropic state in the Bloch vector representation read \cite{Bloch-criterion}:
\begin{eqnarray}
\rho_{\mathrm{W}} & = & \frac{1}{N^2} \mathds{1} \otimes\mathds{1} + \frac{1}{4}\sum_{\mu=1}^{N^2-1} \frac{2(N\phi-1)}{N(N^2-1)} \lambda_{\mu} \otimes \lambda_{\mu} \; , \label{state-W}\\
\rho_{\mathrm{ISO}} & = & \frac{1}{N^2} \mathds{1}\otimes \mathds{1} + \frac{1}{4}(\sum_{\mu \in S_1} \frac{2p}{N} \lambda_{\mu}\otimes \lambda_{\mu} - \sum_{\nu \in S_2} \frac{2p}{N} \lambda_{\nu} \otimes \lambda_{\nu}) \; , \label{state-ISO}
\end{eqnarray}
where $S_1$ represents the symmetric generators of $\lambda_{\mu}^{\mathrm{T}} = \lambda_{\mu}$, and $S_2$ denotes the skew symmetric generators of $\lambda_{\nu}^{\mathrm{T}} = -\lambda_{\nu}$. The partial transposition operation correlates the Werner state with isotropic states. According to Observation \ref{observ-2}, we may readily find that the parameters in Eqs. (\ref{state-W}, \ref{state-ISO}) satisfies
\begin{equation}
p = \frac{N\phi-1}{N^2-1} \; . \label{W-ISO-relation}
\end{equation}
Eq. (\ref{W-ISO-relation}) tells us that, when considering the separability, only one of the two states need to be taken into account. Before proceeding to the separable decomposition, we first present serval straightforward but interesting results from Eq. (\ref{W-ISO-relation}): 1. The positivity condition $\frac{-1}{N^2-1} \leq p $ of $\rho_{\mathrm{ISO}}$ \cite{Reduction-criterion} implies that $\rho_{\mathrm{W}}$ is entangled when $\phi <0$; 2. The positivity condition $\phi \leq 1$ of $\rho_{\mathrm{W}}$ implies that $\rho_{\mathrm{ISO}}$ is entangled when $\frac{1}{N+1} < p $; 3. If $\rho_{\mathrm{W}}$ is separable at $0 \leq \phi \leq 1$ then $\rho_{\mathrm{ISO}}$ is separable at $\frac{-1}{N^2-1} \leq p \leq \frac{1}{N+1}$.

\noindent{\bf Separable decomposition of the generalized Werner state}

Considering the Werner state with $\mathcal{T}= \frac{2(N\phi-1)}{N(N^2-1)} \mathds{1}$ and $\mathrm{rank}(\mathcal{T}) = N^2-1$, there must be at least $N^2$ Bloch vectors in both $M_{rp}$ and $M_{sp}$, due to the additional constraints $M_r\vec{p} = 0$ and $M_{s}\vec{p} = 0$. Based on Theorem \ref{Theorem-main2}, we may construct $M_{rp}$ and $M_{sp}$ as follows:
\begin{align}
M_{rp} & = M_r D_{p}^{\frac{1}{2}} =
\begin{pmatrix}
\alpha_1 & 0 & \cdots & 0 & 0\\
0 & \alpha_2 & \cdots & 0 & 0\\
\vdots & \vdots & \ddots & \vdots & \vdots\\
0 & 0 & \cdots & \alpha_{N^2-1} &0
\end{pmatrix}
Q \; , \label{Mrp-Werner} \\
M_{sp}& = M_s D_{p}^{\frac{1}{2}} =
\begin{pmatrix}
\beta_1 & 0 & \cdots & 0 & 0\\
0 & \beta_2 & \cdots & 0 & 0\\
\vdots & \vdots & \ddots & \vdots & \vdots\\
0 & 0 & \cdots & \beta_{N^2-1} &0
\end{pmatrix}
Q \; . \label{Msp-Werner}
\end{align}
Here, $D_{p}= \mathrm{diag}\{p_1,\cdots, p_{N^2}\}$; $Q \in \mathrm{SO}(N^2)$ whose elements in last row are $Q_{N^2i} = \sqrt{p_i}$ ensuring $M_r\vec{p} = M_{s}\vec{p} = 0$. Since the singular values of $\mathcal{T}$ are all equal, we may set $\alpha_1 =\cdots = \alpha_{N^2-1} = \alpha$, $\beta_1 = \cdots = \beta_{N^2-1} = \beta$, and $p_1 = \cdots = p_{N^2} = \frac{1}{N^2}$; hence have
\begin{equation}
\frac{1}{N} \vec{r}_i = \alpha (Q_{1i}, Q_{2i},\cdots, Q_{(N^2-1)i})^{\mathrm{T}}\; ,\; \frac{1}{N} \vec{s}_i = \beta (Q_{1i}, Q_{2i},\cdots, Q_{(N^2-1)i})^{\mathrm{T}}\; . \label{RS-set}
\end{equation}
The sets $\{\vec{r}_i| i = 1,\cdots, N^2\}$ and $\{\vec{s}_i|i = 1,\cdots, N^2\}$ form two $N^2$-simplexes (or hypertetrahedron) each of which lies in the $(N^2-1)$-dimensional Bloch vector space of particles $A$ and $B$. The angles between any two of the Bloch vectors fulfil
\begin{equation}
\frac{\vec{r}_i \cdot \vec{r}_j}{\vec{r}_i\cdot\vec{r}_i} = \frac{\vec{s}_i \cdot \vec{s}_j}{\vec{s}_i \cdot \vec{s_i}} = -\frac{1}{N^2-1} \; , \; \forall i\neq j \; . \label{pure-simplex-angle}
\end{equation}
Equation (\ref{pure-simplex-angle}) agrees with the requirement for pure state: angle $\theta$ between any two pure states must satisfy (see the Eq. (12) of Ref. \cite{Geo-two-qubit})
\begin{equation}
 -\frac{1}{N-1} \leq \cos \theta  \leq 1\;. \label{pure-Bloch-angle}
\end{equation}
As being true for qubit and numerically verified for qutrit systems, we are tempted to make the following conjecture:
\begin{conjecture}
There exists an $N^2$-simplex with circumradius $\left[\frac{2(N-1)}{N}\right]^{\frac{1}{2}}$, which fits into the convex hull of the $(N^2-1)$-dimensional Bloch vector space of $N$-dimensional mixed states.  \label{Conj-1}
\end{conjecture}

\noindent{\bf The separable decomposition for maximum value of $\frac{2(N\phi-1)}{N(N^2-1)}$}

Conjecture \ref{Conj-1} leads to a solution to the open problem of finding separable decompositions of all separable Werner states in any dimension \cite{S-D-Wener}. Inputting (\ref{RS-set}) to constraints for Bloch vectors, i.e. $|\vec{r}_i|^2, |\vec{s}_i|^2 \leq \frac{2(N-1)}{N}$ (the equality holds for pure state), we have
\begin{equation}
|\vec{r}_i|^2 = N^2 \alpha^2 \sum_{j=1}^{N^2-1} Q_{ji}^2\leq \frac{2(N-1)}{N} \; , \; |\vec{s}_i|^2 = N^2\beta^2 \sum_{j=1}^{N^2-1} Q_{ji}^2\leq \frac{2(N-1)}{N} \; . \label{RS-norm}
\end{equation}
Because $\sum_{j=1}^{N^2-1} Q_{ji}^2= \frac{N^2-1}{N^2}$, Eq. (\ref{RS-norm}) leads to
\begin{equation}
\alpha^2 \leq \frac{2}{N(N+1)} \; , \; \beta^2 \leq \frac{2}{N(N+1)} \; . \label{R-S-1}
\end{equation}
Inputting (\ref{Mrp-Werner}) and (\ref{Msp-Werner}) into $\mathcal{T}= M_{rp}M_{sp}^{\mathrm{T}}$ we have
\begin{equation}
\tau = \left|\frac{2(N\phi-1)}{N(N^2-1)} \right| =\alpha\beta\; . \label{TRS-1}
\end{equation}
Combining of Eqs. (\ref{R-S-1}) and (\ref{TRS-1}) leads to
\begin{equation}
\left|\frac{2(N\phi-1)}{N(N^2-1)} \right| \leq \frac{2}{N(N+1)} \Rightarrow  \frac{2}{N}-1 \leq \phi \leq 1\; .
\end{equation}
The value of $\phi=1$ ($\mathcal{T}_{ii} = \frac{2}{N(N+1)}$) for $\rho_{W}$ has the decomposition of two $N^2$-simplexes in the $N^2-1$ dimensional Bloch vector spaces of particles $A$ and $B$, i.e.
\begin{equation}
\vec{r}_i = \vec{s}_i = \sqrt{\frac{2N}{N+1}} (Q_{1i},\cdots, Q_{(N^2-1)i})^{\mathrm{T}} \; , \; i \in \{1,\cdots, N^2\}\; .
\end{equation}

\noindent{\bf The separable decomposition for minimum value of $\frac{2(N\phi-1)}{N(N^2-1)}$}

If $\mathcal{T}$ is separable when $\phi=1$, and decomposes as
\begin{eqnarray}
\mathcal{T} = \frac{2}{N(N+1)}\mathds{1} = M_{rp}M_{rp}^{\mathrm{T}}\;  \label{r-positive}
\end{eqnarray}
with $M_{rp} =M_{sp}= \{\sqrt{p_1}\,\vec{r}_1, \sqrt{p_2}\, \vec{r}_2, \cdots \}$ and $\vec{r}_i$ being Bloch vectors for pure state, then for $\phi=\frac{2}{N}-1$, $\mathcal{T}$ shall be written as
\begin{eqnarray}
\mathcal{T} = -\frac{2}{N(N+1)}\mathds{1} = -M_{rp}M_{rp}^{\mathrm{T}} = M_{\overline{rp}}M_{rp}^{\mathrm{T}} \; , \label{r-minus-positive}
\end{eqnarray}
where $M_{\overline{rp}} = \{\sqrt{p_1}(-\vec{r}_1), \sqrt{p_2}(-\vec{r}_2), \cdots\}$. If $\vec{r}_i$ in Eq. (\ref{r-positive}) are Bloch vectors of pure state, $-\vec{r}_i$ in Eq. (\ref{r-minus-positive}) can not be physical Bloch vectors for pure state according to Eq. (\ref{pure-Bloch-angle})
\begin{equation}
\frac{(-\vec{r}_i )\cdot \vec{r}_i}{|\vec{r}_i|^2} = -1 \ngeq -\frac{1}{N-1} \; ,
\end{equation}
except for the qubit case of $N=2$, where Bloch vectors of density matrix form a three dimensional ball. Therefore, the lower limit of $\phi$ is not $\frac{2}{N}-1$ except for the qubit case.

Now, suppose one of the two particles having Bloch vectors satisfying Observation \ref{observ-3}, i.e. $|\vec{r}_i|^2 \leq \frac{2}{N(N-1)}$ (or $|\vec{s}_i|^2 \leq \frac{2}{N(N-1)}$, but not both), by the procedure of Eq. (\ref{RS-norm}) to Eq. (\ref{TRS-1}), we have
\begin{equation}
\tau = \left|\frac{2(N\phi-1)}{N(N^2-1)} \right| = \alpha\beta \leq \frac{2}{N(N^2-1)} \Rightarrow  0 \leq \phi \leq \frac{1}{N}\; .
\end{equation}
Therefore the separable decomposition for $\phi=0$ ($\mathcal{T}_{ii} = -\frac{2}{N(N^2-1)}$) reads
\begin{equation}
\vec{r}_i = -N\alpha (Q_{1i}, Q_{2i},\cdots, Q_{(N^2-1)i})^{\mathrm{T}}\; ,\; \vec{s}_i = N\beta (Q_{1i}, Q_{2i},\cdots, Q_{(N^2-1)i})^{\mathrm{T}}\; , \label{small-big}
\end{equation}
where $\alpha^2 = \frac{2}{N(N-1)(N^2-1)}$ and $\beta^2 = \frac{2}{N(N+1)}$. The separable decomposition of Eq. (\ref{small-big}) corresponds to two $N^2$-simplexes: a smaller one composed with $\{\vec{r}_i\}$ and a larger one composed with $\{\vec{s}_i\}$. The smaller one satisfies reflection symmetry: because it lies in the Ball of $|\vec{r}\,|^2 \leq \frac{2}{N(N-1)}$, both $\vec{r}_i$ and $-\vec{r}_i$ are Bloch vectors of density matrices.

\section{Discussion}

We have presented an applicable and operational necessary and sufficient criterion for the separability of bipartite mixed state. The criterion is exhibited in a finite set of inequalities relating the correlation matrix to the Bloch vectors of the quantum states of subsystems, which is shown to be complete by exploring the multiplicative Horn's problem. These inequalities may be treated as trade-off relations between the quantumness of the constituent parts, balanced by the correlation matrix. A state is separable if the decomposition can be performed within the convex hulls of the Bloch vectors of subsystems. As an illustration, separable decompositions for generalized Werner state and isotropic state are achieved in according to the new scheme. The proposed criterion sets up a geometric boundary in between the separability and entanglement for compound system, and provides a new perspective on the nonlocal nature of entanglement.

\section*{Acknowledgements}

\noindent This work was supported in part by the Ministry of Science and Technology of the Peoples' Republic of China(2015CB856703); by the Strategic Priority Research Program of the Chinese Academy of Sciences(XDB23030100); and by the National Natural Science Foundation of China(NSFC) under the grants 11375200 and 11635009.

\newpage
\setcounter{figure}{0}
\renewcommand{\thefigure}{S\arabic{figure}}
\setcounter{equation}{0}
\renewcommand\theequation{S\arabic{equation}}
\setcounter{theorem}{0}
\renewcommand{\thetheorem}{S\arabic{theorem}}
\setcounter{observation}{0}
\renewcommand{\theobservation}{S\arabic{observation}}
\setcounter{lemma}{0}
\renewcommand{\thelemma}{S\arabic{lemma}}
\setcounter{section}{0}
\renewcommand{\thesection}{S\arabic{section}}

\appendix{\bf \LARGE Appendix}

\section{Bloch representation of quantum states}

A quantum state may be represented in form of  Bloch vectors \cite{S-Bloch-vector}
\begin{eqnarray}
\rho = \frac{1}{N}\mathds{1} + \frac{1}{2}  \sum_{\mu=1}^{N^2-1} r_{\mu} \lambda_{\mu} = \frac{1}{N} \mathds{1} + \frac{1}{2} \vec{r}\cdot \vec{\lambda} \; ,
\end{eqnarray}
where the real coefficients $r_{\mu} = \langle \lambda_{\mu} \rangle = \mathrm{Tr}[\rho\lambda_{\mu}]$, and $\lambda_{\mu}$ are the $N^2-1$ traceless generators of SU($N$) group. The $N^2-1$ dimensional real vectors $\vec{r}$ is the Bloch vector representation of the density matrix $\rho$. The normalization of the density matrices $\mathrm{Tr}[\rho^2]\leq 1$ imposes $|\vec{r}\, |\leq \sqrt{2(N-1)/N}$ where the vector norm is defined as $|\vec{r}\, | \equiv \sqrt{\vec{r}\cdot \vec{r}}$. The norm may be regarded as the mixedness (or quantumness) of the quantum state, as $|\vec{r}\, |=0$ corresponds to the completely mixed states while $|\vec{r}\, | = \sqrt{2(N-1)/N}$ corresponds to the pure states.
To ensure the positivity of the density operators, there are further constraints on the Bloch vectors $\vec{r}$ \cite{S-Positivity-Bloch-a, S-Positivity-Bloch-b}. The whole space of the Bloch vectors $\vec{r}$ for quantum states (i.e., the density matrices satisfy the positivity and normalization conditions) forms convex hull in $N^2-1$ dimensional real space, whose circumscribed sphere and inscribed sphere have the radii of \cite{S-Bloch-criterion}
\begin{eqnarray}
R_{+} = \sqrt{\frac{2(N-1)}{N}}\; , \;R_{-} = \sqrt{\frac{2}{N(N-1)}}\; , \label{S-In-out-balls}
\end{eqnarray}
repectively.

An arbitrary bipartite state in the Bloch vector form is
\begin{eqnarray}
\rho_{AB} & = & \frac{1}{NM}  \mathds{1} \otimes \mathds{1} + \frac{1}{2M} \vec{a}\cdot \vec{\lambda}\otimes \mathds{1} + \frac{1}{2N} \mathds{1} \otimes \vec{b} \cdot \vec{\sigma} + \frac{1}{4} \sum_{\mu=1}^{N^2-1}\sum_{\nu=1}^{M^2-1} \mathcal{T}_{\mu\nu} \lambda_{\mu} \otimes \sigma_{\nu} \; , \label{S-bipartite-bloch}
\end{eqnarray}
where $a_{\mu} = \mathrm{Tr}[\rho_{AB} (\lambda_{\mu} \otimes \mathds{1})]$, $b_{\nu} = \mathrm{Tr}[ \rho_{AB} (\mathds{1} \otimes \sigma_{\nu})]$, the correlation matrix $\mathcal{T}_{\mu\nu} = \mathrm{Tr}[\rho_{AB}(\lambda_{\mu} \otimes \lambda_{\nu})]$, $\sigma_{\nu}$ are the generators of SU($M$) and $A$ and $B$ are $N$ and $M$ dimensional subsystems. It could be transformed into the normal form
\begin{equation}
\rho_{AB} \to \widetilde{\rho}_{AB} = \frac{1}{NM} \mathds{1} \otimes \mathds{1} + \frac{1}{4} \sum_{\mu,\nu} \widetilde{\mathcal{T}}_{\mu\nu} \lambda_{\mu} \otimes \sigma_{\nu} \; ,
\end{equation}
where the normal form has the same separability as the original state. Hereafter all the quantum state $\rho_{AB}$ we considered are assumed to be in their normal form, i.e.
\begin{equation}
\rho_{AB} = \frac{1}{NM} \mathds{1} \otimes \mathds{1} + \frac{1}{4} \sum_{\mu,\nu}  \mathcal{T}_{\mu\nu} \lambda_{\mu} \otimes \sigma_{\nu} \; .
\end{equation}
And we have the following observation
\begin{observation}
Let $\vec{r}_i$ and $\vec{s}_j$ be Bloch vectors of density matrices and $\vec{p} = (p_1,p_2,\cdots, p_L)^{\mathrm{T}}$, we may define two matrices $M_{rp} \equiv M_{r}D_{p}^{\frac{1}{2}}$ and $M_{sp} \equiv M_{s}D_{p}^{\frac{1}{2}}$, where $M_r = \{\vec{r}_1, \vec{r}_2, \cdots, \vec{r}_L\}$, $M_s = \{ \vec{s}_1, \vec{s}_2, \cdots, \vec{s}_L\}$, and $D_{p} = \mathrm{diag}\{p_1, p_2, \cdots, p_L\}$ with $0< p_i\leq 1$, $\sum_{i=1}^L p_i = 1$. The state $\rho_{AB}$ is separable if and only if there exist a number $L$ such that $\mathcal{T}= M_{rp} M_{sp}^{\,\mathrm{T}}$ with $M_r \vec{p} = 0$ and $M_s\vec{p} = 0$.
\label{S-Observation-Corr-Mat}
\end{observation}

\section{The multiplicative Horn's problem for real square matrices} \label{S-section-B}

Let $\mathbb{R}^{n\downarrow}$, $\mathbb{R}_+^{n\downarrow}$, and $\mathbb{R}_{+}^{*n\downarrow}$ denote the sets of non-increasing sequences of real numbers, nonnegative real numbers, and strictly positive real numbers respectively. A set $\alpha = \{\alpha_1,\cdots, \alpha_n\} \in \mathbb{R}^{n\downarrow}$ means that $\forall i\in \{1,\cdots,n\}$, $\alpha_i\geq \alpha_{i+1}$ and $\alpha_i \in \mathbb{R}$. $\alpha \leq \beta$ for two sets $\alpha, \beta \in \mathbb{R}^{n\downarrow}$ means that $\forall i \in \{1,\cdots, n\}$, $\alpha_i\leq \beta_i$. $M_n(\mathbb{R})$ and $M_{m\times n}(\mathbb{R})$ denote the sets of real matrices of dimensions $n$ by $n$ and $m$ by $n$ respectively.

\begin{theorem}
$\alpha, \beta, \gamma \in \mathbb{R}_+^{n\downarrow}$ occurs as singular values of $n$ by $n$ real matrices $A$, $B$, and $C = AB$ if and only if
\begin{eqnarray}
\prod_{k\in K} \gamma_k \leq \prod_{i\in I} \alpha_i \cdot \prod_{j\in J} \beta_j
\end{eqnarray}
for all $(I,J,K) \in \bigcup_{r=1}^{n-1} T_r^n$. \label{S-Theorem-triple}
\end{theorem}

We prove this theorem using the method of Ref. \cite{S-Non-invertible}. To proceed the proof we first present the following 7 Lemmas.
\begin{lemma}
$\alpha, \beta, \gamma \in \mathbb{R}^{n\downarrow}$ occurs as eigenvalues of Hermitian $n$ by $n$ matrices $A$, $B$, and $C$ with $C = A+B$ if and only if
\begin{eqnarray}
\sum_{i=1}^n \gamma_i = \sum_{i=1}^n \alpha_i + \sum_{i=1}^n \beta_i \; , \label{S-Horn-sum-1} \\
\sum_{k\in K} \gamma_k \leq \sum_{i\in I} \alpha_i + \sum_{j\in J} \beta_j \; , \label{S-Horn-sum-2}
\end{eqnarray}
hold for every $(I,J,K) \in \bigcup_{r=1}^{n-1} T_r^n$. \label{S-theorem-horn}
\end{lemma}
This is the Horn's conjecture \cite{S-Horn-conjecture} and has been proved in \cite{S-Klyachko-1, S-Knutson-Tao-1}, see \cite{S-Review-Fulton} for a review on this subject where a systematical definition of $T_r^n$ may also be found.

\begin{lemma}
Let $\alpha_i$, $\beta_i$, and $\gamma_i$ be given real numbers, $1\leq i\leq n$. The following conditions are equivalent:
(1) there exist Hermitian matrices $A$, $B$, and $C$ with eigenvalues $\alpha_i$, $\beta_i$, and $\gamma_i$ and sum $C = A+B$; (2) there exist real matrices $X$, $Y$, and $Z$ with singular values $e^{\alpha_i}$, $e^{\beta_i}$, and $e^{\gamma_i}$ and product $ XY = Z$. \label{S-lemma-equivalent-real}
\end{lemma}
This is a direct corollary of Theorem 4.2 in  \cite{S-Real-product}.

\begin{lemma}
For sequences $\lambda = (\lambda_1,\cdots,\lambda_n)$, $\mu = (\mu_1,\cdots,\mu_n)$, $\nu = (\nu_1,\cdots,\nu_n)$ in $\mathbb{R}_{+}^{*n\downarrow}$, there exist matrices $A$, $B$, $C$ $\in$ $M_n(\mathbb{R})$ such that $C = AB $ and having singular values of $\lambda$, $\mu$, and $\nu$, respectively, if and only if
\begin{eqnarray}
&& \prod_{i=1}^n \nu_i =  \prod_{i=1}^n \lambda_i\mu_i \\
&& \prod_{k\in K} \nu_k \leq \prod_{i\in I}\lambda_i \prod_{j\in J}\mu_j\; .
\end{eqnarray}
hold for all $(\lambda,\mu, \nu) \in \bigcup_{r=1}^{n-1} T_{r}^n$. \label{S-lemma-real-product}
\end{lemma}

\noindent{\bf Proof:} From Lemma \ref{S-theorem-horn}, we have that the condition 1 of Lemma \ref{S-lemma-equivalent-real} is equivalent to equations (\ref{S-Horn-sum-1},\ref{S-Horn-sum-2}) which may be reexpressed as
\begin{eqnarray}
\exp(\sum_{i=1}^n \gamma_i) & = & \exp(\sum_{i=1}^n \alpha_i + \sum_{i=1}^n \beta_i) \; , \\
\exp(\sum_{k\in K} \gamma_k ) & \leq & \exp(\sum_{i\in I} \alpha_i + \sum_{j\in J} \beta_j)
\end{eqnarray}
hold for every $(I,J,K) \in \bigcup_{r=1}^{n-1} T_r^n$, where $\alpha$, $\beta$, $\gamma$ in $\mathbb{R}^{n\downarrow} $are eigenvalues of three Hermitian matrices with the third the sum of the first two. Equivalently, we have
\begin{eqnarray}
& & \prod_{i=1}^n e^{\gamma_i}  = \prod_{i=1}^n e^{\alpha_i} e^{\beta_i}\\
& & \prod_{k\in K} e^{\gamma_k} \leq \prod_{i\in I} e^{\alpha_i} \prod_{j\in J} e^{\beta_j}
\end{eqnarray}
hold for every $(I,J,K) \in \bigcup_{r=1}^{n-1}  T_r^n$. According to Lemma \ref{S-lemma-equivalent-real} the condition 2 is equivalent to condition 1 and therefore equivalent to
\begin{eqnarray}
& & \prod_{i=1}^n \nu_i  = \prod_{i=1}^n \lambda_i \mu_i\\
& & \prod_{k\in K} \nu_k \leq \prod_{i\in I} \lambda_i \prod_{j\in J} \mu_j
\end{eqnarray}
hold for all $(I,J,K) \in \bigcup_{r=1}^{n-1} T_r^n$, where $\lambda_i = e^{\alpha_i}$, $\mu_i = e^{\beta_i}$, and $\nu_i = e^{\gamma_i}$. Q.E.D.

For $\lambda,\mu \in \mathbb{R}_+^{n\downarrow}$, define two sets $K_{\lambda,\mu}$ and $\widetilde{K}_{\lambda,\mu}$, where
\begin{eqnarray}
K_{\lambda,\mu}  \equiv  \left\{\nu \in \mathbb{R}_+^{n\downarrow} \mid \nu = \mathrm{Singular\ values\ of}\ \mathrm{diag}\{\lambda\}\ U\ \mathrm{diag}\{\mu\}, \ U\in O(n) \right\} \; ,\hspace{0.7cm} \\
\widetilde{K}_{\lambda,\mu}   \equiv  \left\{\nu \in \mathbb{R}_+^{n\downarrow} \mid \forall (I,J,K) \in \bigcup_{r=1}^{n-1} T_r^n,\prod_{i=1}^n \nu_i =  \prod_{i=1}^n \lambda_i\mu_i, \prod_{k\in K} \nu_k \leq \prod_{i\in I}\lambda_i \prod_{j\in J}\mu_j \right\} \; .
\end{eqnarray}

\begin{lemma}
If $\lambda,\mu \in \mathbb{R}_+^{*n\downarrow}$, then $K_{\lambda,\mu} = \widetilde{K}_{\lambda,\mu}$. \label{S-Lemma-positive}
\end{lemma}
\noindent{\bf Proof:} This is just a different formulation of Lemma \ref{S-lemma-real-product}.

First, if $\nu \in K_{\lambda,\mu}$ then $\nu$ are the singular values of a matrix $C=AB$ where $A$, $B$ have the singular values of $\lambda,\mu$. According to Lemma \ref{S-lemma-real-product}, $\nu \in \widetilde{K}_{\lambda,\mu}$, that is $K_{\lambda,\mu} \subseteq \widetilde{K}_{\lambda,\mu}$.

Second, if $\nu \in \widetilde{K}_{\lambda,\mu}$ where $\lambda,\mu \in \mathbb{R}_+^{*n\downarrow}$, then according to Lemma \ref{S-lemma-real-product} there exist real matrices $C=AB$ where $A$, $B$ are real matrices and have the singular values of $\lambda,\mu$. The singular value decomposition $A = U_a \mathrm{diag}\{\lambda\} V_a^{\mathrm{T}}$, $B = U_b \mathrm{diag}\{\lambda\} V_b^{\mathrm{T}}$, where $U_{a,b},V_{a,b}$ are real orthogonal matrices, tells that $\nu \in K_{\lambda,\mu}$ and thus $\widetilde{K}_{\lambda,\mu} \subseteq K_{\lambda,\mu}$.

Therefore, we have $K_{\lambda,\mu} = \widetilde{K}_{\lambda,\mu}$. Q.E.D.

\begin{lemma}
Let $A,B \in M_n(\mathbb{R})$ and let $C = AB$, and the singular values denote as $\{\sigma_i\} \in \mathbb{R}_+^{n\downarrow}$. Then for every $(I,J,K)\in \bigcup_{r=0}^n T_r^n$ the inequalities
\begin{eqnarray}
\sum_{k\in K} \log \sigma_k(C) \leq \sum_{i\in I} \log \sigma_i(A) + \sum_{j\in J} \log \sigma_j(B)
\end{eqnarray}
holds, with $-\infty$ allowed for the values of $\log$. \label{S-Lemma-infty}
\end{lemma}

\noindent{\bf Proof:} (See Theorem 3.5 of \cite{S-Non-invertible}) Apply polar decomposition to $A$ and $B$ we have $A= U(\sqrt{A^{\dag}A}\,)$ and $B = (\sqrt{BB^{\dag}}\,) V$ where $U,V$ are real orthogonal matrices. Since $C = U (\sqrt{A^{\dag}A}\,) (\sqrt{BB^{\dag}}\,) V$, we get that $\sigma_i(A) = \sigma_i(\sqrt{A^{\dag}A}\,)$, $\sigma_j(B) = \sigma_j(\sqrt{BB^{\dag}}\,)$, and $\sigma_k(C) = \sigma_k(\sqrt{A^{\dag}A} \sqrt{BB^{\dag}}\,)$. Thus, without loss of generality we assume $A$ and $B$ are positive semidefinite. Let $\varepsilon_1,\varepsilon_2 >0$ and let $C(\varepsilon_1,\varepsilon_2) = (A+\varepsilon_1 \mathds{1})(B+\varepsilon_2 \mathds{1})$, we have
\begin{eqnarray}
\sum_{k\in K} \log[ \sigma_k(C(\varepsilon_1,\varepsilon_2))] \leq \sum_{i\in I} \log[\varepsilon_1 + \sigma_i(A)] + \sum_{j\in J} \log[\varepsilon_2 + \sigma_j(B)] \; .
\end{eqnarray}
By letting $\varepsilon_1,\varepsilon_2 \rightarrow 0$, we have $\sigma_k(C(\varepsilon_1,\varepsilon_2)) \rightarrow \sigma_k(C)$, and the right hand side of the inequality has possible values of $-\infty$. Q.E.D.

Define the complement of $(I,J,K)$ in $\{1,\cdots, n\}$ as $(I^c,J^c,K^c)$ where $I\cup I^c = \{1,\cdots, n\}$ and $I\cap I^c = \emptyset$ and similarly for $J^c$ and $K^c$.
\begin{lemma}
Let $\alpha',\alpha'', \beta', \beta'',\gamma', \gamma'' \in \mathbb{R}^{n\downarrow}$, $\alpha'\leq \alpha''$, $\beta'\leq \beta''$, $\gamma' \leq \gamma''$, and $\forall (I,J,K) \in \bigcup_{r=0}^n T^n_r$ the inequalities
\begin{eqnarray}
\sum_{k\in K} \gamma'_k & \leq & \sum_{i\in I} \alpha_i'' + \sum_{j\in J} \beta_j'' \; , \\
\sum_{k\in K^c} \gamma''_k & \geq & \sum_{i\in I^c} \alpha_i' + \sum_{j\in J^c} \beta_j'\; ,
\end{eqnarray}
hold, then there exist $\alpha' \leq \alpha \leq \alpha''$, $\beta' \leq \beta \leq \beta''$, $\gamma' \leq \gamma \leq \gamma''$ that
\begin{eqnarray}
\sum_{i=1}^n \gamma_i & = & \sum_{i=1}^n \alpha_i + \sum_{j=1}^n \beta_j \; , \label{S-interpolation-horn-1} \\
\sum_{k\in K} \gamma_k & \leq & \sum_{i\in I} \alpha_i + \sum_{j\in J} \beta_j \; .\label{S-interpolation-horn-2}
\end{eqnarray}
hold for all $(I,J,K) \in \bigcup_{r=1}^{n-1} T_r^{n}$. \label{S-Lemma-interpolation}
\end{lemma}
Lemma \ref{S-Lemma-interpolation} comes from the Proposition 2.1 of \cite{S-Non-invertible} and Proposition 3.2 of \cite{S-WSLi}.

Define two sets
\begin{eqnarray}
\widetilde{K}^{\leq}_{\lambda,\mu} & = & \left\{ \nu\in \mathbb{R}_+^{n\downarrow}| \forall (I,J,K) \in \bigcup_{r=0}^n T_r^n, \; \prod_{k\in K} \nu_k \leq \prod_{i\in I} \lambda_i \prod_{j\in J} \mu_j \right\}\; , \\
\widetilde{K}^{\geq}_{\lambda,\mu} & = & \left\{ \nu\in \mathbb{R}_+^{n\downarrow}| \forall (I,J,K) \in \bigcup_{r=0}^n T_r^n, \; \prod_{k\in K^c} \nu_k \geq \prod_{i\in I^c} \lambda_i \prod_{j\in J^c} \mu_j \right\} \; .
\end{eqnarray}
Note that $\widetilde{K}_{\lambda,\mu} = \widetilde{K}_{\lambda,\mu}^{\leq}\cap \widetilde{K}_{\lambda,\mu}^{\geq}$, that is the elements both belong to $\widetilde{K}_{\lambda,\mu}^{\leq}$ and $\widetilde{K}_{\lambda,\mu}^{\geq}$ take up the whole set of $\widetilde{K}_{\lambda,\mu}$. Considering the case of $r=n$ where all $I,J,K = \{1,\cdots, n\}$, we have $\prod_{i=1}^n \nu_i \leq \prod_{i=1}^n \lambda_i \prod_{i=1}^n \mu_i$ from $\widetilde{K}_{\lambda,\mu}^{\leq}$. When $r=0$ we have $\prod_{i=1}^n \nu_i \geq \prod_{i=1}^n \lambda_i \prod_{i=1}^n \mu_i$ from $\widetilde{K}_{\lambda,\mu}^{\geq}$. So the values in $\widetilde{K}_{\lambda,\mu}^{\leq}\cap \widetilde{K}_{\lambda,\mu}^{\geq}$ have $\prod_{i=1}^n \nu_i = \prod_{i=1}^n \lambda_i \prod_{i=1}^n \mu_i$.

Now we begin to proof Theorem \ref{S-Theorem-triple}. We need only prove the following lemma
\begin{lemma}
For all $\lambda,\mu \in \mathbb{R}_+^{n\downarrow}$, we have $K_{\lambda,\mu} = \widetilde{K}_{\lambda,\mu}$. \label{S-Lemma-square-final}
\end{lemma}

\noindent{\bf Proof:} First, if $\nu \in K_{\lambda,\mu}$ from Lemma \ref{S-Lemma-infty} we have $\nu \in \widetilde{K}_{\lambda,\mu}$ and $  K_{\lambda,\mu} \subseteq  \widetilde{K}_{\lambda,\mu}$.

Second, if $\nu \in \widetilde{K}_{\lambda,\mu}$, for an arbitrary small positive real number $\varepsilon$, it is clear that, $\nu+\varepsilon \in \widetilde{K}^{\geq}_{\lambda,\mu}$, similarly $\nu \in \widetilde{K}^{\leq}_{\lambda+\varepsilon,\mu+\varepsilon}$. There exist $\delta = \delta(\varepsilon)$ satisfying $0 <\delta < \varepsilon$ such that $\nu + \varepsilon \in \widetilde{K}^{\geq}_{\lambda+\delta,\mu+\delta}$ and $\nu + \delta \in  \widetilde{K}^{\leq}_{\lambda+\varepsilon,\mu+\varepsilon}$.
Thus there exist $\nu_k$ that both of the following two groups of inequalities
\begin{eqnarray}
\sum_{k\in K} \log(\nu_{k} + \delta) \leq \sum_{i\in I} \log(\lambda_i + \varepsilon) + \sum_{j\in J} \log(\mu_j + \varepsilon) \\
\sum_{k\in K^c} \log(\nu_{k} + \varepsilon) \geq \sum_{i\in I^c} \log(\lambda_i + \delta) + \sum_{j\in J^c} \log(\mu_j + \delta)
\end{eqnarray}
hold for $(I,J,K) \in \bigcup_{r=0}^n T_r^n$. Thus according to Lemma \ref{S-Lemma-interpolation} we have
\begin{eqnarray}
\log(\lambda + \delta) \leq \alpha \leq \log(\lambda+\varepsilon) \\
\log(\mu + \delta) \leq \beta \leq \log(\mu + \varepsilon) \\
\log(\nu + \delta) \leq \gamma \leq \log(\nu + \varepsilon)
\end{eqnarray}
where $\sum_{i=1}^n\gamma_i = \sum_{i=1}^n \alpha_i + \sum_{i=1}^n \beta_i$ and $\sum_{k\in K}\gamma_k \leq \sum_{i\in I} \alpha_i + \sum_{j\in J} \beta_j$ for all $(I,J,K) \in \bigcup_{r=1}^{n-1} T_r^n$. Here the inequalities are all assumed to be established componentwise following the definition in the beginning of Sec. \ref{S-section-B}.
Letting $\lambda_{\varepsilon} = e^{\alpha}$, $\mu_{\varepsilon} = e^{\beta}$, and $\nu_{\varepsilon} = e^{\gamma}$, we have
\begin{eqnarray}
\lambda + \delta(\varepsilon) \leq \lambda_{\varepsilon} \leq \lambda + \varepsilon \; , \;
\mu + \delta(\varepsilon) \leq \mu_{\varepsilon} \leq \mu + \varepsilon \; , \;
\nu + \delta(\varepsilon) \leq \nu_{\varepsilon} \leq \nu + \varepsilon \; . \label{S-twoside-varepsilon}
\end{eqnarray}
As $\nu_{\varepsilon} \in \widetilde{K}_{\lambda_{\varepsilon},\mu_{\varepsilon}}$, Lemma \ref{S-Lemma-positive} gives that $\nu_{\varepsilon} \in K_{\lambda_{\varepsilon},\mu_{\varepsilon}}$, so there is an real orthogonal matrix $U_{\varepsilon}$ so that the singular values of $\mathrm{diag}(\lambda_{\varepsilon}) U_{\varepsilon} \mathrm{diag}(\mu_{\varepsilon})$ are precisely $\nu_{\varepsilon}$.
Choosing a sequence $\varepsilon(k)$ tending to zero as $k \to \infty$, so that $U_{\varepsilon(k)}$ converge to a real orthogonal matrix $U$ as $k \to \infty$. We have that the singular values of $\mathrm{diag}(\lambda) U \mathrm{diag}(\mu)$ are precisely $\nu$ based on Eq. (\ref{S-twoside-varepsilon}). Thus $\nu \in K_{\lambda,\mu}$ and $\widetilde{K}_{\lambda,\mu} \subseteq K_{\lambda,\mu}$.

From the above discussions we have $K_{\lambda,\mu} = \widetilde{K}_{\lambda,\mu}$.  Q.E.D.

The Lemma \ref{S-Lemma-square-final} is equivalent to Theorem \ref{S-Theorem-triple}.

\section{Examples}

\subsection{Comparing with the PPT criterion}

The partial transposition is an operation on the density matrix that makes the transposition only on one of the composed particles, i.e.,
\begin{eqnarray}
( A\otimes B)^{\mathrm{T}_{\mathrm{B}}} = A\otimes B^{\mathrm{T}} \; .
\end{eqnarray}
So a partial transposition of a separable state
\begin{eqnarray}
\rho_{AB} = \sum_{k}p_k \rho^{(A)}_k \otimes \rho^{(B)}_k \; , \;\sum_k p_k=1\; , \;  p_k>0
\end{eqnarray}
takes the form of
\begin{eqnarray}
\rho_{AB}' = \rho_{AB}^{\mathrm{T}_{\mathrm{B}}}  = \sum_{k} p_k \rho^{(A)}_k \otimes \rho'^{(B)}_k\; ,\; \rho'^{(B)}_k = \rho^{(B)\mathrm{T}}_k\; ,
\end{eqnarray}
which is also a separable quantum state and thus positive definite. Therefore, the positive partial transposition criterion is a necessary condition for the quantum state to be separable: if a quantum state is separable then the quantum state and its partial transposition are positive definite.

In our method, the partial transposition on partite $B$ of quantum state $\rho_{AB}$ is
\begin{eqnarray}
(\rho_{AB})^{\mathrm{T}_{\mathrm{B}}} & = & \frac{1}{N^2} \mathds{1} \otimes \mathds{1} + \frac{1}{4} \sum_{\mu,\nu=1}^{N^2-1}t_{\mu\nu}\lambda_{\mu} \otimes \lambda_{\nu}^{\mathrm{T}} \nonumber \\
& = & \frac{1}{N^2} \mathds{1} \otimes \mathds{1} + \frac{1}{4} \sum_{\mu,\nu=1}^{N^2-1} t'_{\mu\nu} \lambda_{\mu} \otimes \lambda_{\nu}\; ,
\end{eqnarray}
where $t'_{\mu\nu}=-t_{\mu\nu}$ for the columns that $\lambda_{\nu}^{\mathrm{T}} = -\lambda_{\nu}$. If $\rho_{AB}$ is separable that $\mathcal{T} = M_rM_s^{\mathrm{T}}$, then
\begin{eqnarray}
(\rho_{AB})^{\mathrm{T}_{\mathrm{B}}} & = & \frac{1}{N^2} \mathds{1} \otimes \mathds{1} + \frac{1}{4} \sum_{k=1}^{n}p_k \vec{r}_k \cdot \vec{\lambda} \otimes \vec{s}_k\,\!' \cdot \vec{\lambda}\; ,
\end{eqnarray}
where $\vec{r}_k\cdot \vec{\lambda} = \sum_{\mu} (r_k)_{\mu} \lambda_{\mu}$, $\vec{s}_k\,\!' \cdot \vec{\lambda} = \sum_{\mu}(s'_k)_{\nu}\lambda_{\nu}$, and  $(s_k')_{\nu} = -(s_k)_{\nu}$ for the skew symmetric $\lambda_{\nu}$. Because the transposition of the density matrix of one particle quantum state is also a density matrix of one particle quantum state, so $(\rho_{AB})^{\mathrm{T}_{\mathrm{B}}}$ is also a separable state that is positive definite, i.e., PPT is necessary.

The sufficient part of the PPT criterion is: if $(\rho_{AB})^{\mathrm{T}_{\mathrm{B}}}$ is positive definite ($\rho_{AB}$ is positive definite by definition) then $\rho_{AB}$ is separable. This statement is not true for general bipartite states. Here we show that it is true for qubit-qubit systems. In the normal forms, the two-qubit quantum state may be represented as $\rho_{AB} = \frac{1}{4}x_{\mu\nu}D_{\mu\nu}$ \cite{S-two-qubit-bloch}, where $D_{\mu\nu} = \sigma_{\mu} \otimes \sigma_{\nu}$, $\sigma_{0}=\mathds{1}$, and $\sigma_{1,2,3}$ are Pauli matrices. The relation of $X=x_{\mu\nu}$ ($\mu,\nu \in \{0,1,2,3\}$) with $\mathcal{T} = t_{\mu\nu}$ ($\mu,\nu \in \{1,2,3\}$) is
\begin{eqnarray}
X =
\begin{pmatrix}
  1 & 0 & 0 & 0 \\
  0 & t_{11} & t_{12} & t_{13} \\
  0 & t_{21} & t_{22} & t_{23} \\
  0 & t_{13} & t_{23} & t_{33}
\end{pmatrix} =
\begin{pmatrix}
  1 & 0 \\
  0 & \mathcal{T}
\end{pmatrix} \; .
\end{eqnarray}
The positivity condition of $\rho_{AB}$ requires (see Eq. (54) in \cite{S-two-qubit-bloch})
\begin{eqnarray}
4 - ||X||^2 \geq 0 \; ,\\ -2 \mathrm{det} \mathcal{T} -(||X||^2-2) \geq 0 \; , \\
-8\mathrm{det}\mathcal{T} + (||X||^2-2)^2 -4(\tau_2^2\tau_3^2 + \tau_3^2\tau_1^2 + \tau_1^2\tau_2^2) \geq 0 \; .
\end{eqnarray}
where $||X||^2 = \mathrm{Tr}[X^{\dag}X]$, $\tau_i$ are the singular values of $\mathcal{T}$. After the partial transposition, $\rho_{AB}^{\mathrm{T}_{\mathrm{B}}}$ has the $X'$ in following form
\begin{eqnarray}
X' =
\begin{pmatrix}
  1 & 0 & 0 & 0 \\
  0 & t_{11} & -t_{12} & t_{13} \\
  0 & t_{21} & -t_{22} & t_{23} \\
  0 & t_{13} & -t_{23} & t_{33}
\end{pmatrix} \; , \; \mathcal{T}\,' =
\begin{pmatrix}
 t_{11} & -t_{12} & t_{13} \\
 t_{21} & -t_{22} & t_{23} \\
 t_{13} & -t_{23} & t_{33}
\end{pmatrix} \; .
\end{eqnarray}
Therefore the positivity condition of $\rho_{AB}^{\mathrm{T}_{\mathrm{B}}}$ requires
\begin{eqnarray}
4 - ||X||^2 \geq 0 \; ,\\ 2 \mathrm{det} \mathcal{T} -(||X||^2-2) \geq 0 \; , \\
8\mathrm{det}\mathcal{T} + (||X||^2-2)^2 -4(\tau_2^2\tau_3^2 + \tau_3^2\tau_1^2 + \tau_1^2\tau_2^2) \geq 0 \; ,
\end{eqnarray}
where we have used the fact $\mathrm{det}[\mathcal{T}\,'] = -\mathrm{det}[\mathcal{T}]$ and the singular values of $\mathcal{T}\,'$ and $\mathcal{T}$ are the same. The two group of positivity inequalities should be satisfied simultaneously, so we have
\begin{eqnarray}
\tau_1^2+\tau_2^2+\tau_3^2 \leq 3 \; , \\
\pm 2\tau_1\tau_2\tau_3 -(\tau_1^2 + \tau_2^2 + \tau_3^2 -1) \geq 0 \; , \\
\pm 8 \tau_1\tau_2\tau_3 + (\tau_1^2 + \tau_2^2 + \tau_3^2 -1)^2 -4(\tau_2^2\tau_3^2 + \tau_3^2\tau_1^2 + \tau_1^2\tau_2^2) \geq 0 \; ,
\end{eqnarray}
which be reduced to
\begin{eqnarray}
\tau_1^2+\tau_2^2+\tau_3^2 \leq 3 \; , \\
\tau_1^2 + \tau_2^2 + \tau_3^2 +  2\tau_1\tau_2\tau_3 \leq 1 \; , \\
(\tau_1-\tau_2-\tau_3-1) (\tau_1+\tau_2-\tau_3+1) (\tau_1-\tau_2+\tau_3+1) (\tau_1+\tau_2+\tau_3-1) \geq 0 \; .
\end{eqnarray}
Further reductions shows that $0\leq \tau_1 + \tau_2 + \tau_3 \leq 1$, and according Corollary 2 we have the qubit-qubit system is separable. Therefore the PPT critrion is necessary and sufficient for qubit-qubit systems.

\subsection{ The decomposition of the generalized Werner state}

Here we present a numerical result for the decomposition of qutrit Werner state
\begin{equation}
\rho_{\mathrm{W}} = \frac{1}{9}\mathds{1} \otimes \mathds{1} + \frac{1}{4} \sum_{\mu = 1}^{8} \frac{2(3\phi -1)}{3\times 8} \lambda_{\mu} \otimes \lambda_{\mu} \; .
\end{equation}
For  $\phi = 1$ we have
\begin{equation}
\rho_{\mathrm{W}} = \frac{1}{9}\mathds{1} \otimes \mathds{1} + \frac{1}{4} \sum_{\mu = 1}^{8} \frac{1}{6} \lambda_{\mu} \otimes \lambda_{\mu} \; .
\end{equation}
The Bloch vectors are
\begin{align}
\vec{r}_i = (\frac{3}{2})^{\frac{1}{2}} (Q_{1i}, Q_{2i},\cdots, Q_{8i})^{\mathrm{T}}\; , \\
\vec{s}_i = (\frac{3}{2})^{\frac{1}{2}} (Q_{1i}, Q_{2i},\cdots, Q_{8i})^{\mathrm{T}}\; ,
\end{align}
where $i\in \{1, \cdots, 9\}$. The constraints of $Q_{ij}$ ($Q_{9i} = \frac{1}{3}$) are
\begin{align}
\sum_{k=1}^8 Q_{ki}Q_{kj} =
\left\{
\begin{array}{ll}
\frac{8}{9} & i=j \\ \\
\frac{-1}{9} & i\neq j
\end{array} \right. \; , \label{S-SO-orth} \\
-\frac{4}{9} +  \frac{1}{2} \left(\frac{3}{2}\right)^{\frac{3}{2}} d_{\mu\nu\rho}Q_{\mu i} Q_{\nu i} Q_{\rho i} = 0 \; , \label{S-Positive-3}
\end{align}
where Eq. (\ref{S-Positive-3}) ensures that the density matrices of $\rho_i^{(A)}$ and  $\rho_i^{(B)}$ are all positive semidefinite \cite{S-Positivity-Bloch-b}. Numerical analysis of Eqs. (\ref{S-SO-orth}, \ref{S-Positive-3}) could be evaluated using math softwares (there exist subroutines in Mathematica or MATLAB).

\subsection{ Pure Bell state}

For Bell state $|\psi\rangle_{AB} =\frac{1}{\sqrt{2}}( |00\rangle + |11\rangle )$, its density matrix is
\begin{eqnarray}
\rho_{AB} = \frac{1}{2}
\begin{pmatrix}
1 & 0 & 0 & 1 \\
0 & 0 & 0 & 0 \\
0 & 0 & 0 & 0 \\
1 & 0 & 0 & 1
\end{pmatrix} = \frac{1}{4}(\mathds{1}\otimes \mathds{1} + \sigma_x\otimes \sigma_{x} - \sigma_y\otimes \sigma_y + \sigma_z\otimes \sigma_z) \; .
\end{eqnarray}
The correlation matrix $\mathcal{T} = \mathrm{diag}\{1,-1,1\}$ and has the singular values $\tau_1=\tau_2=\tau_3=1$. Corollary 1 leads to $(\tau_1+\tau_2+\tau_3)^2 = (\sum_{i}p_i|\vec{r}_i|^2) ( \sum_{j}p_j|\vec{s}_j|^2)$, where no Bloch vectors $\vec{r}_i$ and $\vec{s}_j$ of qubit systems satisfy the relation ($\tau_1+\tau_2+\tau_3=3$ while $\sum_i p_i = 1$, $|\vec{r}_i|^2\leq 1$ and $|\vec{s}_i|^2\leq 1$).

\subsection{P-zero state}

Consider the following bipartite qubit state \cite{S-Bloch-criterion}
\begin{eqnarray}
\rho_{\pm} = p|\psi^{\pm}\rangle\langle \psi^{\pm}| + (1-p)|00\rangle\langle 00|\; ,
\end{eqnarray}
where $0\leq p\leq 1$ and $|\psi^{\pm}\rangle = \frac{1}{\sqrt{2}}(|01\rangle \pm |10\rangle)$. If $p=0$ then $\rho_{\pm}$ is a separable state. If $p\neq 0$, we may apply the procedure of \cite{S-normal-forms} to transform $\rho_{\pm}$ to their normal forms. For the case of $\rho_+$, after $N$ steps of transformations, we have
\begin{eqnarray}
\widetilde{\rho}_+ = \frac{1}{2}
\begin{pmatrix}
\frac{2-2p}{N+1-Np} & 0 & 0 & 0 \\
0 & \frac{N-1-(N-2)p}{N+1-Np} & \sqrt{\frac{N-1-(N-2)p}{N+1-Np}} & 0 \\
0 & \sqrt{\frac{N-1-(N-2)p}{N+1-Np}} & 1 & 0 \\
0 & 0 & 0 & 0
\end{pmatrix}\; .
\end{eqnarray}
We see that its normal form is $\lim_{N\to \infty} \widetilde{\rho}_+ = |\psi\rangle\langle\psi|$, where $|\psi\rangle = \frac{1}{\sqrt{2}}(|01\rangle + |10\rangle)$. Similar process applies to $\rho_-$, therefore $\rho_{\pm}$ is separable only when $p=0$.

\end{document}